\documentclass[8.5pt,twoside,twocolumn]{article}
\usepackage[super,sort&compress,comma]{natbib}
\usepackage{mhchem}
\usepackage{stfloats}
\usepackage{times,mathptmx}
\usepackage{amsmath}
\usepackage{sectsty}
\usepackage{balance}

\usepackage{graphicx} 
\usepackage{lastpage}
\usepackage[format=plain,justification=raggedright,singlelinecheck=false,font=small,labelfont=bf,labelsep=space]{caption}
\usepackage{fancyhdr}
\begin{document}

\thispagestyle{plain}
\fancypagestyle{plain}{
\renewcommand{\headrulewidth}{1pt}}
\renewcommand{\thefootnote}{\fnsymbol{footnote}}
\renewcommand\footnoterule{\vspace*{1pt}%
\hrule width 3.4in height 0.4pt \vspace*{5pt}}
\setcounter{secnumdepth}{5}

\makeatletter
\def\subsubsection{\@startsection{subsubsection}{3}{10pt}{-1.25ex plus -1ex minus -.1ex}{0ex plus 0ex}{\normalsize\bf}}
\def\paragraph{\@startsection{paragraph}{4}{10pt}{-1.25ex plus -1ex minus -.1ex}{0ex plus 0ex}{\normalsize\textit}}
\renewcommand\@biblabel[1]{#1}
\renewcommand\@makefntext[1]%
{\noindent\makebox[0pt][r]{\@thefnmark\,}#1}
\makeatother
\renewcommand{\figurename}{\small{Fig.}~}
\sectionfont{\large}
\subsectionfont{\normalsize}

\fancyfoot{}
\fancyfoot[RO]{\footnotesize{\sffamily{1--\pageref{LastPage} ~\textbar  \hspace{2pt}\thepage}}}
\fancyhead{}
\renewcommand{\headrulewidth}{1pt}
\renewcommand{\footrulewidth}{1pt}
\setlength{\arrayrulewidth}{1pt}
\setlength{\columnsep}{6.5mm}
\setlength\bibsep{1pt}

\twocolumn[
  \begin{@twocolumnfalse}
\noindent\LARGE{\textbf{Gender Bias in Nobel Prizes}}
\vspace{0.6cm}

\noindent\large{\textbf{Per Lunnemann$^a$, Mogens H. Jensen$^{b}$, and Liselotte Jauffred$^{b}$}}\\ \\
\footnotesize{\textit{$^{a}$ Blackwood Seven, Livj{\ae}gergade 17B, 2. Floor
2100 Copenhagen, Denmark. $^{b}$ The Niels Bohr Institute, University of Copenhagen, Blegdamsvej 17, Denmark.; E-mail: jauffred@nbi.dk.}}


\vspace{0.6cm}

\noindent 
\normalsize{Strikingly few Nobel laureates within medicine, natural and social sciences are women. Although it is obvious that there are fewer women researchers within these fields, does this gender ratio still fully account for the low number of female Nobel laureates? We examine whether women are awarded the Nobel Prizes less often than the gender ratio suggests. Based on historical data across four scientific fields and a Bayesian hierarchical model, we quantify any possible bias. The model reveals, with exceedingly large confidence, that indeed women are strongly under-represented among Nobel laureates across all disciplines examined.
}
\vspace{0.5cm}
 \end{@twocolumnfalse}
  ]

%
%
%
\begin{figure*} [b]
\begin{center}
  \includegraphics[width=17cm]{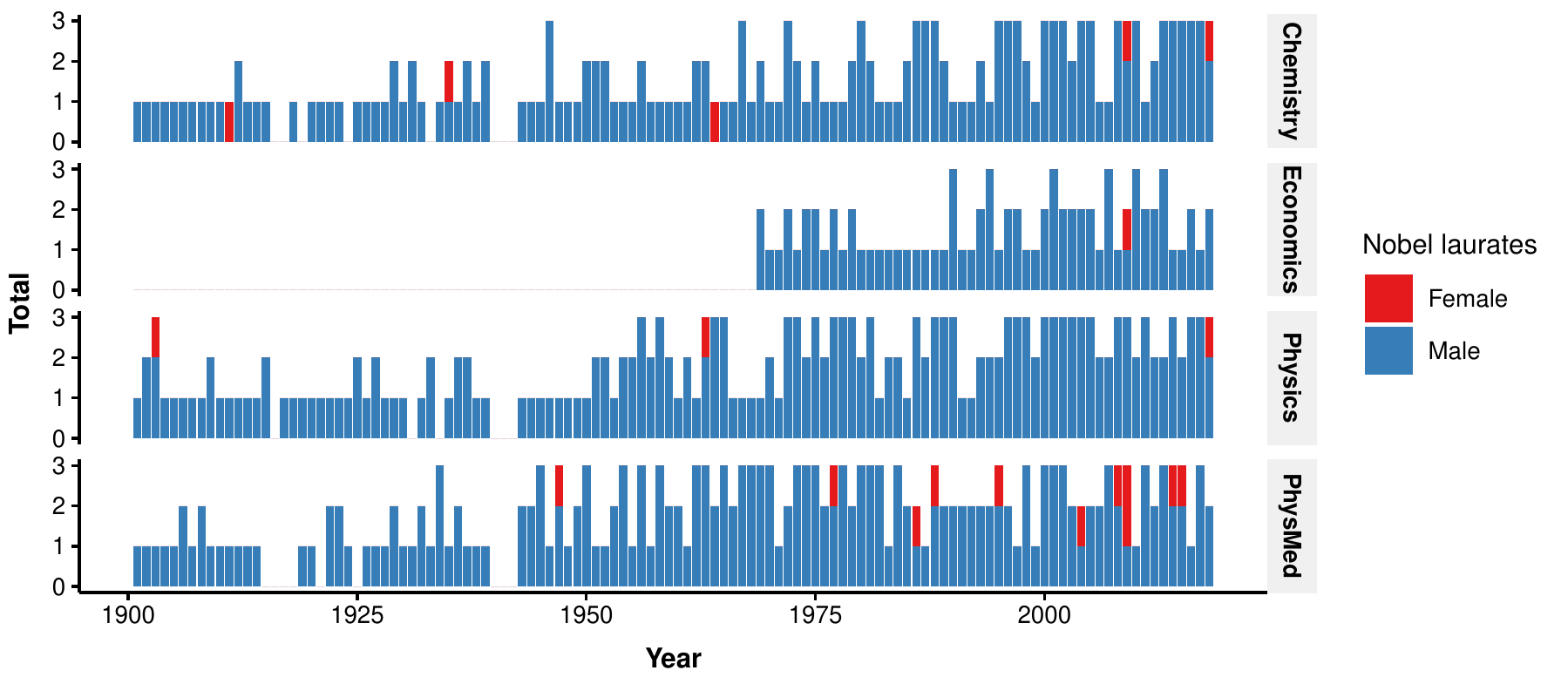}
  \captionof{figure}{Gender distribution of Nobel Prizes. Bar plot of the scientific Nobel Prizes from 1901 to 2018.\cite{NP}}
\label{fig:barplot}
\end{center}
\end{figure*}
\noindent
This year Prof. D. Strickland received the Nobel Prize in Physics as the first woman in 55 years. From 1901 to 2018, the Nobel Prize in Physics has been awarded 112 times to 209 different candidates; among these are only two more women; namely M. Curie in 1903 and M. Goeppert Mayer in 1963. Women have historically occupied much fewer positions in academia than men, so naturally one would expect more male Nobel laureates than female. In case this was the only important factor, we would expect the Nobel awards to follow a binomial distribution with a probability given by the gender ratio. For instance, if there is 10\% women within a field, we expect \emph{ceteris paribus} a 10\% chance that a woman is awarded the Nobel Prize. 
But does the gender ratio truly account for the few female Nobel laureates?

To investigate this, we compared the gender ratio of Nobel laureates in Physics; Chemistry, Economics, Physiology, and Medicine to the relevant gender ratios among scientists in the field. We use the gender distribution of senior faculty members in the US as proxy for a worldwide distribution and observe that women are awarded the Nobel Prize far less often than the faculty gender ratios suggest. More specifically, we find the probability that the distribution of Nobel Prizes is not favoring men, to be less than 4\% for within all of the investigated fields. 

\section*{Results}
Since the first Nobel Prizes were awarded in 1901 there has been 688 Nobel laureates within the fields of Chemistry, Economics, Physics, and Medicine; among these are only 21 women, see Fig. \ref{fig:barplot}. Among the Nobel laureates of economics there is one woman; namely Prof. E. Ostrom (2009) which corresponds to 2\%. In Medicine, 12 women have been awarded over the years which 6\% of the laureates. It is obvious that these differences reflect, to some extent, the gender ratios within the field.   
However, the gender distribution of faculty members evolves and for every instance in time, the gender distribution among senior faculty members is different than junior faculty members. As the average age of Nobel laureates is 55 years\cite{NP}, we assume that the Nobel laureates are sampled from a gender distribution of senior faculty members. Moreover, Nobel laureates did their ground breaking findings a few decades prior to the award (the average is 15 years\cite{NP}). To account for this, we assume that today's Nobel laureates are sampled from senior faculty members $\delta$ years ago.

We examined the fraction of female faculty members relative to all faculty members which we denoted gender ratios, $r$. We used the gender ratios of senior faculty members at university departments in the US as a proxy for a global distribution. The data were retrieved from the National Science Foundation \cite{NSF1973-2010} and covers the period from 1973 to 2010. For completeness, we fitted with a logistic function and extrapolated the data back in time to cover the entire period of Nobel awards from 1901 to 2010, see Fig. \ref{fig:PriorPosterior}. In the data, both Chemistry and Physics are gathered under Physical sciences. Hence, we used this gender ratio for both the Physics and Chemistry Nobel Prizes. Furthermore, for the Nobel Prize in Economics we used the gender ratio of senior faculty members from Social sciences. Most probably, this leads to a slight overestimation of the bias within economics, since economics may have a smaller gender ratio than the overall ratio within Social sciences. 
\begin{center}
  \includegraphics[width=8cm]{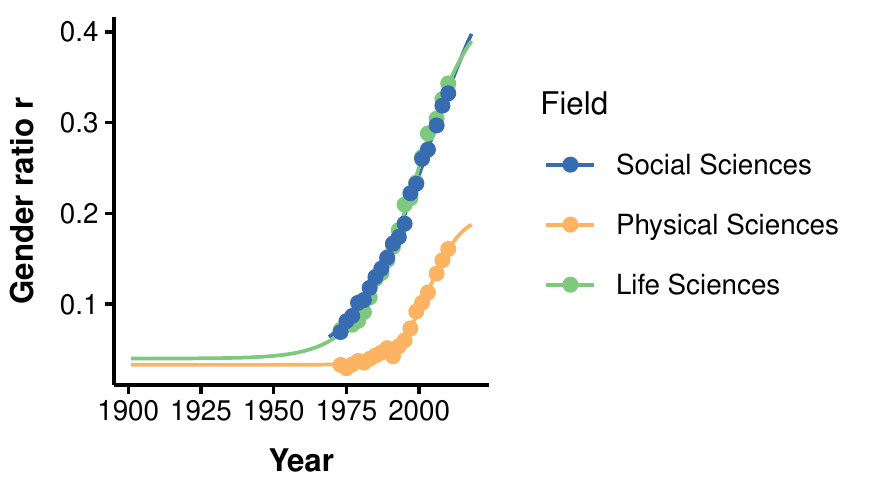}
  \captionof{figure}{Gender ratio, $r$, defined as the number of women relative to all faculty members, versus years: data (points) from the National Science Foundation \cite{NSF1973-2010} and fit of a logistic function (line).}
    \label{fig:PriorPosterior}
\end{center}

We use a hierarchical model to quantify possible gender bias in the awarding of Nobel Prizes using Bayesian inference through Hamiltonian Monte Carlo sampling, see Methods section. The gender bias is described by the parameter $\alpha$ and when $\alpha < 1$ ($\alpha > 1$), women are awarded the Nobel Prize less (more) often than the gender ratio suggests. The sampled prior and posterior probability density distributions, $p(\alpha | \mathbf{r},\delta)$, is illustrated in Fig. \ref{fig:alpha}, for a lag of $\delta=10$ and ratios $\mathbf{r}$. From the prior distribution (grey) we confirm that we chose a weakly informative prior, allowing values of $\alpha$ both well below and above 1. For all four Nobel Prizes, the posterior distributions shows a significant bias against women with mean values of the posterior probability density $\langle \alpha \rangle < 1$ and a total probability of being larger than unity, $P(\alpha\geq1)=1-\int_0^1 p(\alpha|r,\delta)\mathrm{d}\alpha$, found to be less than a few percents.
To investigate how sensitive the measured bias is to the choice of $\delta$ we repeated the analysis in the range $0\leq\delta\leq20$. For all values of $\delta$, sample values of $\alpha$ were predominantly smaller than unity. This is summarized in Fig. \ref{fig:marginal}, which shows the probability of $\alpha$ being larger than 1, $P(\alpha\geq1)$, versus delay, $\delta$.
\begin{center}
  \includegraphics[width=8cm]{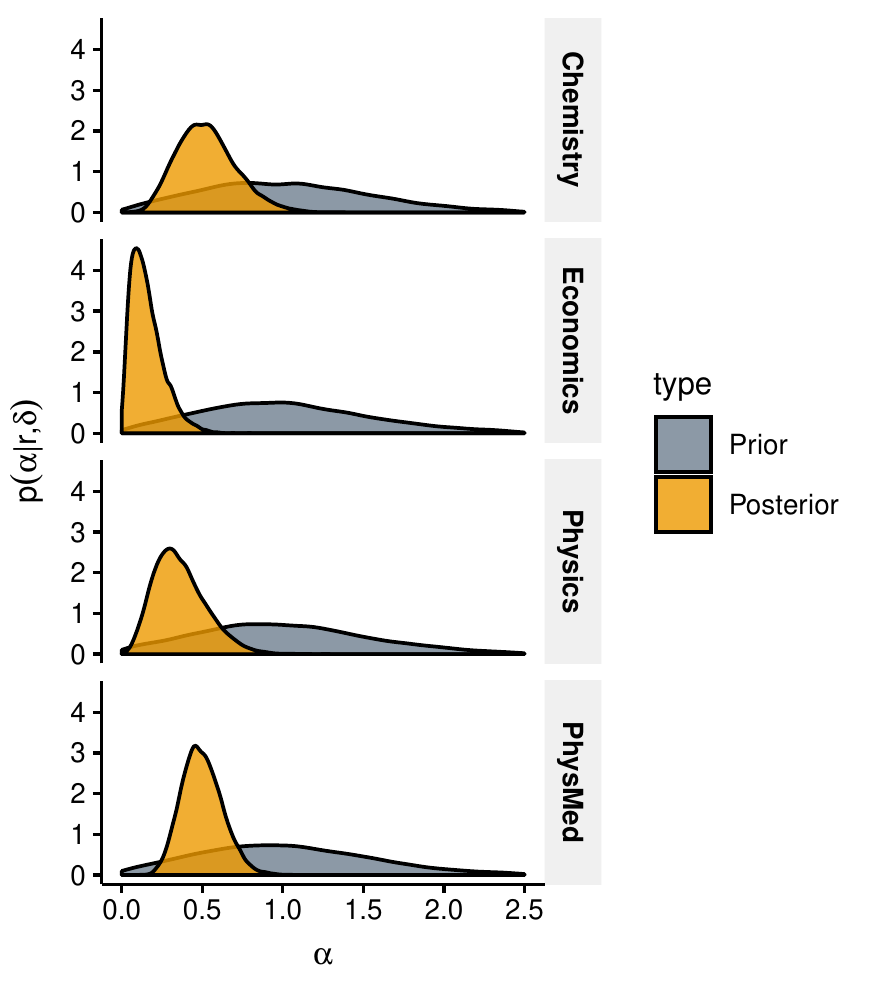}
  \captionof{figure}{Prior (grey) and posterior (orange) probability density of $\alpha$, for $\delta=10$. The prior distribution was set giving a mean of 1, see Eq. \eqref{eq:alphaPrior} and Eq. \eqref{eq:muPrior}. Values of $\alpha$ less (more) than unity signify fewer (more) Nobel Prizes awarded women than the gender ratio suggests.}
    \label{fig:alpha}
\end{center}

We anticipate that the variations within the different fields, to some extent, reflect the granularity of categories in the historical gender ratios. For instance, for Economics (blue curve) we probably overestimated the bias by comparing with the gender ratio within Social Sciences (where we believe the ratio is larger). In contrast, for Chemistry (red curve) we were likely to underestimate the bias as we collated the prizes with the ratio of Physical Sciences, which includes both Physics and Chemistry. 
Therefore, we do not conclude that Nobel Prizes  for some scientific field have a larger bias, than for others. Regardless of this, we find that the possibility that Nobel laureates are awarded without disfavouring women is less than 4\% for lag times less than 20 years. This firm evidence shows that women are disfavored, i.e., female senior scientists are less likely to be awarded a Nobel Prize than their gender ratio suggests. 
Furthermore, one could argue that the findings are often done early in the career, i.e., before tenure, where gender ratios are more balanced. If this is true, our model underestimates the bias against women. 

\section*{Discussion}
Using a hierarchical Bayesian interference model we found that the gender distribution in Nobel Prizes includes a bias against women with more than $\sim96\%$ probability. Hence, even women that resist the \textit{leaky pipeline} \cite{Pell1996} and become permanent staff members do not have equal chances to become a Nobel laureate.
However, our models do not propose that this bias arises from an unfair evaluation of nominees by the Nobel committees. In contrast, we believe that this divergence occurs at multiple earlier steps in the careers of potential Nobel laureates. This means that there is not an equal possibility for both genders to be nominated for a Nobel Prize. We speculate that there are limitations for women to enter the pool of very well esteemed scientists worthy of a nomination. This hindrance could be related to family life; as female laureates are significantly less likely to be married and/or have children than male laureates.\cite{Charyton2011} Furthermore, there are indications that men in academia are more likely to be provided the resources and support needed for an excellent scientific production.\cite{Xie2003} This suggests that men are more prone to end up in the pool of possible Nobel nominees. Therefore, these results are not only of relevance for future Nobel laureates, but for all future faculty members.
\begin{center}
  \includegraphics[width=8cm]{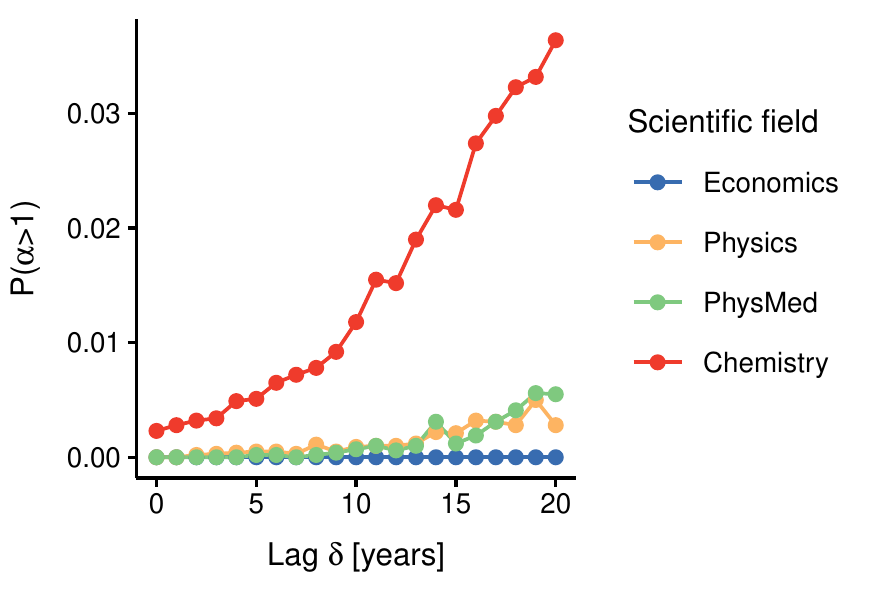}
  \captionof{figure}{Estimated probability of $\alpha$ being more than one, $P(\alpha\geq1)=1-\int_0^1 p(\alpha)\mathrm{d}\alpha$, i.e., the probability that women are favoured versus delay parameter $\delta_i$.}
    \label{fig:marginal}
\end{center}

\section*{Methods}
We used data of faculty members resolved on gender and fields from the National Science Foundation \cite{NSF1973-2010} as a proxy for a global distribution. This data only covers the period from 1973 to 2010, hence, we extrapolated the data with a logistic function to obtain the gender ratios, $r$, for different fields from 1901 to 2010. We note that the average age for Nobel laureates is 55 years and the findings, worthy of a Nobel Prize, are on average done 15 years earlier.\cite{NP} While we do not have access to the number of female and male faculty members, resolved by age, we define a lag time, $\delta$. With this, we presume that the relevant research originate from senior faculty members $\delta$ years before.   

We use a hierarchical Bayesian inference model and Hamiltonian Monte Carlo sampling.\cite{Carpenter2017} We model the number of women laureates, $f_i$, in year $i$ as a stochastic binomial variable:
\begin{equation}
    f_i \sim B(N_i,\theta_i),
\end{equation}
where $B$ is the binomial distribution. $N_i$ and $\theta_i$ are the number of Nobel Prizes awarded and the corresponding success probability, i.e., the probability of a women being awarded, in year $i$, respectively. 
In the case of no bias, we expect $\theta_i$ to be equal to the gender ratio, $r_{i-\delta}$ some $\delta$ years earlier. In order to quantify any bias we model the success probability, $\theta$ as
\begin{equation}
\theta_i  =  \alpha r_{i-\delta},
\end{equation}  
where $\alpha$ is a positive, time independent, stochastic variable fulfilling  $0\leq \alpha \leq 1/r_i$ for all years $i$. Here, $\alpha$ is a bias parameter, such that when $\alpha =1$, women are awarded the Nobel Prize exactly as often as the gender ratio suggests. We use a hierarchical structure for the variable $\alpha$, assuming, for each scientific field, $f$, that the mean of $\alpha_f$ is drawn from a stochastic (hyper) variable $\mu$. Hence, we assume some similarity between the four different $\alpha_f$'s. We use
\begin{align}
\alpha_f &\sim N(\mu, 0.35)\label{eq:alphaPrior}\\
\mu & \sim \Gamma(5,5),\label{eq:muPrior}
\end{align}
where $N$ and $\Gamma$ are the normal and $\Gamma$-distribution, respectively.
Hence, we choose a weakly informative prior distribution for $\alpha_f$ with a mean of 1 and standard deviation of roughly 0.57, see Fig. \ref{fig:alpha}. We further note that the results were found significantly robust on the choice of the hyper parameter $\mu$ and on the standard deviation of the normal distribution, Eq. \eqref{eq:alphaPrior}.

\providecommand{\latin}[1]{#1}
\providecommand*\mcitethebibliography{\thebibliography}
\csname @ifundefined\endcsname{endmcitethebibliography}
  {\let\endmcitethebibliography\endthebibliography}{}

\end{document}